\documentclass[11pt,letterpaper]{article}

\usepackage{graphicx}

\begin{document}

\def\beq{\begin{equation}}
\def\eeq{\end{equation}}
\def\bea{\begin{eqnarray}}
\def\eea{\end{eqnarray}}

\def\a{\alpha}
\def\r{\rho}
\def\s{\sigma}
\def\m{\mu}
\def\n{\nu}
\def\k{\kappa}
\def\g{\gamma}
\def\L{\Lambda}
\def\D{\Delta}
\def\la{\langle}
\def\ra{\rangle}
\def\o{\omega}
\def\d{\delta}
\def\p{\partial}
\def\Se{$S_E$ }
\def\Sa{$S_{\rm atmo}$ }

\def\tphi{\tilde{\phi}}
\def\tu{\tilde{u}}
\def\hv{\hat{v}}

\def\nab{\nabla}

\def\half{\textstyle{\frac{1}{2}}}
\def\quarter{\textstyle{\frac{1}{4}}}

\begin{center} {\large \bf
Black Hole Thermodynamics
and
Lorentz Symmetry}

\vskip 5mm
\large
{Ted Jacobson$^{*}$\footnote{E-mail:
jacobson@umd.edu} and Aron C. Wall$^{*}$\footnote{E-mail:
aronwall@umd.edu}}

\vskip  0.5 cm
{\centerline{\it Department of Physics}}
{\centerline{\it University of Maryland}} 
{\centerline{\it College Park, MD 20742-4111, USA}} 
\vskip  0.5 cm
\today
\end{center}
\vskip  0.5 cm

\begin{abstract}
Recent developments 
point to a breakdown in the generalized second law
of thermodynamics for theories with Lorentz symmetry violation.
It appears possible to construct a perpetual motion
machine of the second kind in such theories,
using a black hole to catalyze the conversion of heat
to work. Here we describe 
and extend
the arguments leading to
that conclusion. We suggest the 
inference that local Lorentz symmetry may be 
an emergent property of the macroscopic world
with origins in a microscopic second law of causal
horizon thermodynamics. 
\end{abstract}

A black hole interior is hidden from the outside world 
by a causal barrier, and so black holes seemed at first 
to challenge our understanding of the
second law of thermodynamics.  
Entropy-carrying heat could be dumped inside, 
never to be heard from again. 
But it was found 
that thermodynamics 
{\it can} be successfully generalized to black holes once quantum mechanical effects such as Hawking radiation are taken into account.
A black hole satisfies the laws of thermodynamics with a
particular temperature $T_H=\hbar\k/2\pi$ and entropy
$S_{BH}=A/4\hbar G$~\cite{Bekenstein:1973ur,Hawking:1974sw}, where $\k$ is the surface
gravity and $A$ the area of the horizon.
Central among these laws is the
generalized second law, according to which the sum of the
black hole entropy and the ordinary entropy outside the black hole
never decreases.

Yet recent developments indicate that this thermodynamic paradise cannot be extended to the case of a Lorentz-violating (LV) theory, whether the violation arises from spontaneous symmetry breaking or a fundamental term in the Lagrangian.  Generically, LV theories have terms causing different fields to propagate at different speeds, and even when such terms are artificially zeroed out they usually still reappear due to renormalization effects.  We will therefore focus on the case of theories where different fields travel at different speeds, bearing in mind that exceptional LV theories may exist where our analysis does not apply.

Lorentz-violating 
theories possessing fields with
different maximum propagation speeds can be coupled to
a metric gravity theory. In black hole solutions of
such theories, different fields would 
typically
see different
event horizons, with different surface gravities and
hence different Hawking temperatures.  Such black
holes would therefore not be in equilibrium, despite
being classically static configurations. Moreover, as
will be explained in what follows, such a system could
apparently be exploited to convert heat to work, in a
cycle that returns the system---insofar as it
observable on the outside---to its original state.
Ignoring the black hole interior, this would violate
Kelvin's heat engine version of the second law of
thermodynamics, which states the impossibility of such
a cycle, also known as a perpetual motion machine of
the second kind.

The second law has various formulations: (i) Kelvin's
heat engine version,
(ii) Clausius' heat flow version
which states the impossibility of making heat flow
from a colder to a hotter reservoir with no change in
the surroundings, and (iii) Clausius' entropy version,
stating that the total entropy cannot decrease.
Clausius and Kelvin presumed that all parts of the
thermodynamic system are accessible, and under that
presumption the different versions of the second law
are equivalent. However, since the
state of a black hole interior can never influence the
outside world, it is natural to adapt these versions
of the second law by  excluding the black hole
interior from the ``system" and its ``surroundings".
With this interpretation, the heat versions (i) and
(ii) of the second law can be violated in a LV theory.
Also the entropy outside the black hole can be
decreased by a process in which the  external
appearance of the black hole is unchanged, so this
``outside entropy" version of the second law can be
violated. By unitarity, the entropy version for the
{\it total} entropy on a complete spacelike surface,
including any inside the black hole, remains
unthreatened by LV. As for the {\it generalized
entropy} $S_{\rm outside} + S_{\rm BH}$, it is unclear
in this LV context how to even define the black hole
entropy. By returning the black hole to the same
exterior state, the need to specify $S_{BH}$ is
eliminated.  Violation of the generalized second law is
therefore possible for any choice of $S_{BH}$.

 Dubovsky and Sibiryakov (DS) 
 constructed a perpetual motion
machine in a two-speed LV theory~\cite{Dubovsky:2006vk}. 
A black hole in such a
theory has a higher Hawking temperature for the
``fast" field $B$ than for the ``slow" field $A$. DS
showed that one could therefore arrange for $A$-heat
to flow in and $B$-heat to flow out at a higher
temperature. The particular design required that there
be no interaction between the $A$ and $B$ fields
(although gravitational interaction cannot be
avoided). Several potential flaws in this design for a
perpetual motion machine were analyzed in
Ref.~\cite{Eling:2007qd}, where it was concluded that
these flaws can be evaded by working with a
sufficiently large black hole. A more efficient
machine, based on purely classical operating
principles, was also discussed there.

This classical mechanism
mimics
Penrose's original procedure~\cite{Penrose:1971uk} for
extracting energy from a rotating black hole. He noted
that there are negative energy states in the
ergoregion outside the horizon. (``Energy" refers here
to the globally conserved quantity that exists on
account of the time translation symmetry of the
stationary rotating black hole spacetime.) 
By lowering
a massive particle into such a negative energy state,
one can extract more energy than the particle's rest
mass.
The
extra comes from the rotational energy of the hole. 

A non-rotating black hole also has an ergoregion, but it
lies causally isolated behind the horizon. 
But when the black hole has two horizons the story is different.
There are negative
energy $A$-states lying behind the $A$-horizon yet
outside the $B$-horizon, and these {\it can} be
exploited using $A$-$B$ interactions. 
Ref.~\cite{Eling:2007qd} considered scenarios in which some number of classical $A$ and $B$ particles fall into the ergoregion, either separately or in a bound state.  They then split up into massless $A$ and $B$ particles, the former carrying negative energy into the black hole, and latter carrying positive energy out.  Here we propose a variant of this process, in which an $A$-object is lowered on a $B$-rope past the
$A$-horizon, into the erogregion above the $B$-horizon. 
The rope must descend sufficiently rapidly that the worldline of the
$A$-object remains $A$-timelike.
Then the object is dropped in and the rope is pulled back out.  (This assumes that there exist interactions between the $A$ and $B$ particles capable of producing the tension in the rope, and the bond between the rope and the $A$-object.)  This process
extracts the entire rest mass of the $A$-particle, plus
some of the mass-energy of the black hole, 
as work at
infinity.

To violate the heat version of the second
law, one can 
adjust the process by suspending a $B$-box containing
thermal $A$-radiation on the $B$-rope, and lowering the
box to
the $A$-horizon.
The $A$-heat can then
be discarded in the black hole, thus converting it
into work at infinity, without changing the
exterior appearance of the
black hole nor using up any exterior fuel.
It might seem that
in the usual Lorentz invariant case
heat could also be converted
to work by lowering the heat to the horizon
and dropping it in. However, the presence
of the acceleration
radiation or thermal atmosphere of the
black hole appears to preclude this
possibility~\cite{Unruh:1982ic,Marolf:2003wu}.

One potential flaw in this perpetual motion machine is
the quantum instability in which
positive energy $B$-particles and negative energy
$A$-particles spontaneously
appear in the ergoregion, the former
escaping to infinity and the latter falling deeper
into the hole.
This instability has some finite rate and could be balanced by an influx of energy.  However, one might worry about whether this leads to a compensating entropy increase outside.  
For dimensionless couplings between the $A$ and $B$ fields, this entropy increase should be of order unity per light-crossing time (the same order rate as the Hawking evaporation process).  However, as shown in 
Ref.~\cite{Eling:2007qd}, classical entropy extraction processes can occur at a much quicker rate than this, so this effect can be neglected.

Another potential flaw is that the $B$-box and rope will
emit $A$-\v{C}erenkov radiation as they are withdrawn from
the $A$-ergoregion. Thus it will take extra work to pull them
back up. However, this extra work is not directly related to
the amount of $A$ heat energy extracted, so by proper design
the perpetual motion machine should still operate.

Yet another possibility is that 
perhaps any two-speed LV theory can only have a UV completion
if it violates locality and allows escape of entropy
from the interior of the
black hole~\cite{Dubovsky:2006vk,ArkaniHamed:2007ky}.
This might indeed restore the
validity of the ``exterior second law" (by eliminating the
interior!) but it
would be a very strange theory.
Not only would it be massively non-local, it would 
need to
have many non-localizable low energy states
(since one could extract the
energy from entropy before dumping it into the black hole).
In particular, no \emph{generalized} second law (GSL), in which information inside the horizon may be ignored, would hold.
 
The implication of these thought experiments is that any LV theory in which different fields have different horizons violates the GSL.  But what if the horizons happened to coincide? Even if the fields travel at different speeds in general, the light cones of all fields might be tangent to the same horizon.  In such a theory the particular GSL-violating mechanisms described here would not apply.  This possibility was invoked in Refs.~\cite{Sagi:2007hb} and \cite{Betschart:2008yi} to argue that some LV theories might satisfy the GSL.  However in both of these examples, at least one of the fields is singular on the horizon, so it is not really clear whether these objects are black holes rather than naked singularities, or whether they have thermal Hawking radiation in all fields.  Moreover, the proposed counterexamples of Ref.~\cite{Betschart:2008yi} are not obtained from any theory with specified equations of motion, so it is not yet possible to ask how the horizon entropy changes in response to infalling matter.

Another counterexample to the classical entropy lowering process was proposed in Ref.~\cite{Mukohyama:2009um} in the context of the ghost condensate theory.  It was pointed out that the energy extraction rate via a two-particle classical scattering processes is much less than the rate of energy increase due to accretion of the ghost condensate.  However if the particles fall into the ergoregion in a bound state, or are lowered with a rope as described here, then the kinematical constraints used in 
Ref.~\cite{Mukohyama:2009um} to bound the rate of energy extraction are absent.  We conclude that none of these counterexamples invalidate our conclusion that the GSL can be violated by classical energy extraction in multiple speed theories.

It is not surprising that Lorentz violation would threaten the GSL.  After all, at the quantum level the thermality of the Hawking radiation is closely related to the 
Unruh effect---the fact that the Minkowski vacuum is thermal 
with respect to 
the boost Hamiltonian when restricted to the Rindler wedge. 
This in turn depends crucially on the Lorentz symmetry of the quantum field theory vacuum 
\cite{Sewell,Unruh:1983ac}.  
So there is simply no reason to expect the Hawking radiation to be thermal, and therefore no reason to expect the GSL to hold at the quantum level.   Note that this argument, unlike the others, applies even for LV theories in which the speeds of all fields are the same in the infrared (perhaps as a result of fine-tuning parameters in the action, or supersymmetry \cite{GrootNibbelink:2004za,Bolokhov:2005cj}).

Taken together, all the classical and quantum arguments discussed above strongly suggest that LV and black hole thermodynamics are incompatible.  But we need not infer from this that the only way to keep black hole thermodynamics is to start with a theory that has Lorentz symmetry as a geometrical action on points and fields.
If a microscopic theory has the feature that black hole horizons always satisfy the GSL, Lorentz invariance will always be found on the macroscopic level.

The physical mechanisms that would account for this tendency are 
obscure.  But the proposal indicates some possible features of a quantum gravity theory.  The key concept in black hole thermodynamics is the notion of a causal horizon.  
When a pure state is restricted to the region outside of a horizon, it becomes a mixed state. ÊBut mixed states can define a ``modular flow'', which is a notion of time with respect to which the state is thermal \cite{Connes:1994hv}. Ê
For the Rindler wedge in Minkowski spacetime this modular flow is the Lorentz boost. This suggests that more generally the local modular flow defined by a
causal region could give rise to a local 
Lorentz boost symmetry.  
The GSL would imply relations between  
the modular flows of different
causal regions, and those relationships
might be sufficient to guarantee that 
these regions can be stitched together
into a macroscopic Lorentzian spacetime
with local Lorentz symmetry.  
In any case, the connection between black hole 
thermodynamics and Lorentz symmetry seems
a tantalizing clue hinting at the fundamental nature of quantum spacetime. 


\section*{Acknowledgements}
 This work was 
supported in part by the National Science Foundation under grants
No. PHY-0601800 and PHY-0903572.

\end{document}